\documentclass[twocolumn,showpacs, nofootinbib,aps,superscriptaddress, eqsecnum,prd,prl,notitlepage,showkeys,10pt,reprint]{revtex4-1}

\usepackage{amsmath}
\usepackage{amsfonts}
\usepackage{amssymb}
\usepackage{graphicx}
\usepackage{color}
\usepackage{hyperref}
\usepackage{booktabs}
\usepackage{hyperref}
\usepackage{cleveref}
\usepackage{color}
\usepackage{dcolumn}
\usepackage{epstopdf}

\def\mnras{Mont. Not. Roy. Astr. Soc.}

\def\apjl{Astrop. Jour. Lett.}
\def\aap{Astronomy \& Astrophysics}

\def\jcap{Jour. Cosm. Astrop. Phys.}
\def\apss{Astrop.\&Sp. Sc. }

\Crefname{equation}{Eq.}{Eqs.}
\Crefname{figure}{Fig.}{Figs.}
\Crefname{section}{Sec.}{Secs.}

\usepackage{etoolbox}
\makeatletter
\appto{\appendix}{%
  \@ifstar{\def\theequation@prefix{A.}}%
          {}%
}
\makeatother

\begin{document}

\title{Speeding up the universe using dust with pressure}

\author{Orlando Luongo}	
\email{orlando.luongo@lnf.infn.it}
\affiliation{Istituto Nazionale di Fisica Nucleare, Laboratori Nazionali di Frascati, 00044 Frascati, Italy.}
\affiliation{School of Science and Technology, University of Camerino, I-62032, Camerino, Italy.}
\affiliation{Instituto de Ciencias Nucleares, Universidad Nacional Aut\'onoma de M\'exico, AP 70543, M\'exico DF 04510, Mexico.}

\author{Marco Muccino}
\email{marco.muccino@lnf.infn.it}
\affiliation{Istituto Nazionale di Fisica Nucleare, Laboratori Nazionali di Frascati, 00044 Frascati, Italy.}

\begin{abstract}
We revise the cosmological standard model presuming that matter, i.e. baryons and cold dark matter, exhibits a non-vanishing pressure mimicking the cosmological constant effects. In particular, we propose a scalar field Lagrangian $\mathcal L_1$ for matter with the introduction of a Lagrange multiplier as constraint. We also add a symmetry breaking effective potential accounting for the classical cosmological constant problem, by adding a second Lagrangian $\mathcal{L}_2$. Investigating the Noether current due to the shift symmetry on the scalar field, $\varphi\rightarrow\varphi+c^0$, we show that $\mathcal{L}_1$ turns out to be independent from the scalar field $\varphi$. Further we find that a positive Helmotz free-energy naturally leads to a negative pressure without introducing by hand any dark energy term. To face out the fine-tuning problem, we investigate two phases: before and after transition due to the symmetry breaking. We propose that during transition dark matter cancels out the quantum field vacuum energy  effects. This process leads to a negative and constant pressure whose magnitude is determined by baryons only. The numerical bounds over the pressure and matter densities are in agreement with current observations, alleviating the coincidence problem. Finally assuming a thermal equilibrium between the bath and our effective fluid, we estimate the mass of the dark matter candidate. Our numerical outcomes seem to be compatible with recent predictions on WIMP masses, for fixed spin and temperature. In particular, we predict possible candidates whose masses span in the range $0.5-1.7$ TeV.
\end{abstract}

\maketitle

\section{Introduction}
\label{sec1}

The $\Lambda$CDM concordance paradigm is described by the fewest number of assumptions possible. In particular, the universe is approximated at late times by two fluids: pressureless matter and a cosmological constant $\Lambda$. Both baryonic matter (BM) and cold dark matter (DM) are unable to push the universe to accelerate \cite{2006IJMPD..15.1753C}. Thus, besides dust-like fluids, one needs to include $\Lambda$ to account for the observed speed up. The simplicity of the concordance paradigm turns out to be the strong suit to admit its validity. However, the magnitude of $\Lambda$ predicted by quantum fluctuations of flat space-times leads to a severe fine-tuning problem with the observed value of $\Lambda$. Even considering a curved space-time one cannot remove the problem \cite{1989RvMP...61....1W}. Further, both matter and $\Lambda$ magnitudes are extremely close today, leading to the well-known \textit{coincidence problem} \cite{1995A&A...301..321W,1999PhRvD..60d3501A,2003PhRvD..67h3513C}. Under these aspects the $\Lambda$CDM model seems to be incomplete, whereas from a genuine observational point of view it well adapts to data.

In this work, we revise the cosmological standard model assuming an effective a scalar field $\varphi$ Lagrangian for baryons and cold DM. We require that matter provides a \emph{non-vanishing pressure term} and we wonder whether it can accelerate the universe alone, i.e. without the need of $\Lambda$. To do so, we propose the most general Lagrangian, depending upon a \emph{kinetic term and Lagrange multiplier}, with the inclusion of a potential term due to the vacuum energy cosmological constant, inducing a \emph{phase transition}. During such an early-time phase transition the DM pressure counterbalances the $\Lambda$ pressure, leaving as unique contribute the pressure of baryons\footnote{At the end of transition.}. In particular, the baryonic pressure turns out to be negative to guarantee a positive Helmotz free-energy for the whole system. In this picture, we find a Noether current, coinciding with the entropy density current and providing the Lagrangian to be independent from $\varphi$. We thus write up the thermodynamics associated to the model and we investigate small perturbations, finding the adiabatic and non-adiabatic sound speeds naturally vanish in analogy to the $\Lambda$CDM approach.

Our paradigm candidates as an alternative to the concordance model and predicts the existence of a single fluid, composed of baryons and cold DM with pressure. The fluid cancels out the quantum contribution due to $\Lambda$, driving the universe today with a constant and negative pressure. This process does not set $\Lambda$ to zero, but removes it naturally. This is possible if DM constituents lie on the mass interval $\sim0.5$--$1.7$~TeV. To show this, we relate the predictions of our model to the thermal history of the primordial universe and to the expected DM relic abundance.

The paper is structured as follows. In Sec.~\ref{sec:action}, we propose the effective representation for matter with pressure.
We thus write the equations of motion and we discuss the introduction of the potential term due to the vacuum energy cosmological constant.
In Sec.~\ref{sec:thermodynamics}, we describe the thermodynamics of our matter fluid, which \emph{naturally} suggests an emergent negative pressure, and demonstrate that our Lagrangian does not depend upon $\varphi$.
In Sec.~\ref{pert} we investigate the small perturbations and we find that both the adiabatic and non-adiabatic sound speeds naturally vanish, leading to a constant pressure, in analogy to the $\Lambda$CDM approach.
In Sec.~\ref{vacuumenergy} we closely analyze the role of the effective potential $V^{\rm eff}$. This term induces a first order transition phase during which the quantum field vacuum energy density mutually cancels with the DM pressure.
Soon after the transition the emergent $\Lambda$ is given by the (negative) pressure of baryons.
We show how our mechanism overcomes the fine-tuning and the coincidence issues affecting the $\Lambda$CDM model.
In Sec.~\ref{DMparticle} we relate the predictions of our paradigm to observable quantities. We thus obtain that, almost independently from the spin, the DM mass ranges within the interval: $0.5\lesssim M{\rm c^2/ TeV} \lesssim1.7$.
In Sec.~\ref{predictions} we summarize the main predictions of our model and then in Sec.~\ref{conclusions} we propose conclusions and perspectives of our work.

\section{The effective representation of matter with pressure}\label{sec:action}

We discussed in Sec.~\ref{sec1} that our model pushes up the universe to accelerate by means of matter with pressure. In particular, we want to demonstrate that the DM pressure may counterbalance the effects of $\Lambda$, if a transition phase is involved in our picture. To show that, let us start from a few number of hypotheses that we will use later on, summarized below:
\begin{itemize}
\item[-] there exists only \emph{one fluid}, composed of BM and DM;
\item[-] matter is coupled to $\Lambda$ and the coupling effect cancels the cosmological constant density which does not enter the dynamical equations;
\item[-] the process which cancels the effects of $\Lambda$ is due to a \emph{first order phase transition};
\item[-] the whole  kinetic energy of matter is constrained through a Lagrange multiplier;
\item[-] the thermodynamics of matter \emph{naturally} suggests an emergent negative pressure;
\item[-] the model mimes the $\Lambda$CDM effects, without departing from observations at both late and early stages of universe's evolution.
\end{itemize}

This suggests an effective representation of \emph{dust with pressure} in a curved space-time given by Lagrangian density $\mathcal{L}=\mathcal{L}_1+\mathcal{L}_2$, where
\begin{align}
\label{eq:1}
\mathcal{L}_1 &= K\left(X,\varphi\right) +\lambda Y\left[X,\nu\left(\varphi\right)\right]\,,\\
\label{eq:1bi}
\mathcal{L}_2 &= -V^{\rm eff}\left(X,\varphi\right)\,,
\end{align}
depend upon the scalar field $\varphi$ and its first covariant derivatives\footnote{Higher order derivatives are excluded because of the \textit{Ostrogradski's theorem}: systems characterized by a non-degenerate Lagrangian dependent on time derivatives of higher than the first leads to a linearly unstable Hamiltonian function.} in the form of the standard kinetic term
\begin{equation}
\label{eq:2}
X = \frac{1}{2}g^{\alpha\beta}\nabla_\alpha \varphi \nabla_\beta\varphi\,,
\end{equation}
where $g^{\alpha\beta}$ is the metric tensor and $\nu(\varphi)$ plays the role of the specific inertial mass \cite{1970PhRvD...2.2762S}.

The Lagrangian $\mathcal{L}_1$ represents a dust component with pressure.
It is written in the most generic form without indicating a priori the functional forms of the functions $Y$ and $K$, while the Lagrange multiplier $\lambda$ constraints the kinetic energy with the potential term in $\nu$. The physical motivation behind $\mathcal{L}_1$ supports the idea of BM and DM with pressure \cite{2000ApJ...535L..21M,2006MNRAS.372..136F,2017arXiv170701059S,bet}.
It is important to stress that our fluid consists of BM and DM, so that in principle the Lagrangian $\mathcal{L}_1$ can be written as
\begin{equation}
\label{BM+DM}
\mathcal{L}_1=K_{\rm BM}+ K_{\rm DM} + \lambda \left(Y_{\rm BM} + Y_{\rm DM}\right)\,,
\end{equation}
where $K\equiv K_{\rm BM}+K_{\rm DM}$ and $Y\equiv Y_{\rm DM}+Y_{\rm DM}$.

The Lagrangian $\mathcal{L}_2$ models the coupling with the standard cosmological constant through an interacting potential $V^{\rm eff}$ used to investigate the phase transition. We write down the simplest form of $V^{\rm eff}$ by
\begin{equation}
\label{eq:vint}
V(\varphi,\psi)=V_0+\frac{\chi}{4}\left(\varphi^2-\varphi^2_0\right)^2+\frac{\bar{g}}{2}\varphi^2\psi^2,
\end{equation}
in which the first two terms describe the self-interacting potential, with a dimensionless coupling constant $\chi$, of the scalar field $\varphi$, and the last one the interacting potential, with a dimensionless coupling constant $\bar{g}$, between $\varphi$ and another scalar field $\psi$.
The quantity $V_0$ denotes
the classical off-set, while $\varphi^2_0$ is the value of $\varphi$ at the minimum of its potential without interactions with $\psi$.
We can thus assume that $\psi$ is in thermal equilibrium. In such a case, $\psi^2$ can be replaced through its average in a thermal state. In a thermal state there exists a correspondence between the thermal average state and the temperature. Following \cite{2006ftft.book.....K}, we redefine the coupling constant $\bar{g}$ to account for the proportionality between the thermal average and the temperature, i.e. we have
$$\langle\psi^2\rangle_{T}\propto T^2\,.$$
After some manipulations, we simple have
\begin{equation}
\label{eq:pottemperature}
V^{\rm eff}(X,\varphi)=V_0+\frac{\chi}{4}\left(\varphi^2-\varphi_0^2\right)^2+\frac{\chi}{2}\varphi_0^2\varphi^2
\frac{T^2(X)}{T_{\rm c}^2}\,,
\end{equation}
where $T_{\rm c}=\varphi_0\sqrt{\chi/\bar{g}}$ is the critical temperature, discriminating as transition starts. Before the transition when $T>T_{\rm c}$, the minimum of $V^{\rm eff}$ is located at $\varphi=0$ and the corresponding value is $V_0+\chi \varphi_0^4/4$.
After the transition, when $T<T_{\rm c}$ the minimum is  at $\varphi=\varphi_0$ with a value $V_0$.

From Eqs.~\eqref{eq:1}--\eqref{eq:1bi} we define the action $
S = \int \mathcal{L}\,\sqrt{-g}{\rm d}^4x$, where $g$ is the determinant of $g^{\alpha\beta}$.
Assuming a standard minimal coupling with gravity, from the variation of the action with respect to $\lambda$, $\varphi$ and the metric tensor we obtain a \textit{constraint} and a \textit{dynamical} equation and the energy-momentum tensor respectively (details of calculations are reported in Appendix~\ref{appA})
\begin{align}
\label{eq:no5a}
&\,Y = 0\,, \\
\label{eq:no5b}
&\,\mathcal{L}_\varphi - \nabla_\alpha \left( \mathcal{L}_X \nabla^\alpha\varphi \right) = 0\,,\\
\label{eq:no6}
&\, T_{\alpha\beta}  = \mathcal{L}_X \nabla_\alpha\varphi \nabla_\beta \varphi - \left(K -  V^{\rm eff} \right) g_{\alpha\beta}\,,
\end{align}
where the subscripts label the partial derivatives, so that $\mathcal{L}_X=K_X-V^{\rm eff}_X+\lambda Y_X$ and $\mathcal{L}_\varphi = K_\varphi-V^{\rm eff}_\varphi+\lambda Y_\nu \nu_\varphi$.
For time-like derivatives it holds $X>0$ and, from Eq.~(\ref{eq:2}), we can introduce an effective $4$-velocity
\begin{equation}
\label{eq:no7}
v_\alpha = \frac{\nabla_\alpha\varphi}{\sqrt{2X}}\ ,
\end{equation}
while the $4$-acceleration identically vanishes
\begin{equation}
\label{eq:no8}
a_\beta = \dot{v}_\beta = v_\gamma \nabla^\gamma v_\beta = 0\,,
\end{equation}
where $\dot{y}=v^\alpha\nabla_\alpha y$ is the Lie derivative of $y$ along $v^\alpha$, which is tangent to time-like geodesics.
Using Eq.~(\ref{eq:no7}), the energy-momentum tensor can be written as
\begin{equation}
\label{eq:no10}
T_{\alpha\beta} = 2X \mathcal{L}_X v_\alpha v_\beta - \left(K -  V^{\rm eff} \right) g_{\alpha\beta}\ ,
\end{equation}
which is of the perfect fluid form for an energy density and a pressure, respectively
\begin{align}
\label{eq:no11}
\rho\left(\lambda,X,\varphi\right) =\,& 2X \mathcal{L}_X - \left(K -  V^{\rm eff} \right)\,,\\
\label{eq:no12}
P\left(X,\varphi\right) =\,& K - V^{\rm eff}\,.
\end{align}
Thus, from the above definitions one wonders whether it is possible to fulfill the weak energy conditions
$T_{\alpha\beta}k^\alpha k^\beta \geq0, \rho \geq 0, \rho + P \geq 0$,
where $k^\alpha$ is a time-like vector field.
From the above conditions and the fact that $X>0$, it follows that
\begin{equation}
\label{weaks}
2X\mathcal{L}_X \geq K - V^{\rm eff}\,,\qquad \mathcal{L}_X \geq 0\,.
\end{equation}
The energy-momentum tensor conservation gives
\begin{equation}
\label{eq:no13}
\nabla_\alpha T^{\alpha\beta} = \left[\dot{\rho} + \theta \left(\rho + P\right) \right] v^\beta = 0\ ,
\end{equation}
leading to the energy conservation $
\dot{\rho} + \theta\left(P + \rho \right) = 0$ where we defined the \textit{expansion}
\begin{equation}
\label{eq:no15}
\theta = \nabla_\alpha v^\alpha = \frac{\nabla_\alpha\nabla^\alpha \varphi-X_\varphi}{\sqrt{2X}}\,.
\end{equation}
The energy flux $T^{\alpha\beta}v_\beta=\rho v^\alpha$ always follows time-like geodesics, as for the perfect fluid with no pressure. By means of this position
\begin{equation}
\label{eq:no16}
\eta_\varphi =
2X \left(\mathcal{L}_{XX} X_\varphi + \mathcal{L}_{X\varphi} \right) + \mathcal{L}_X X_\varphi -\mathcal{L}_\varphi\,,
\end{equation}
the constraint in Eq.~(\ref{eq:no5b}) can be written as follows
\begin{equation}
\label{eq:no9b}
\dot{\lambda} = -\frac{1}{2X Y_X} \left[\sqrt{2X}\eta_\varphi + \theta\left(P + \rho \right) \right] \,.
\end{equation}

\noindent Eqs.~(\ref{eq:no5a}) and (\ref{eq:no9b}) represent the equations of motion for a perfect fluid ruled by two first-order ordinary differential equations for the scalar fields $\varphi$ and $\lambda$ \cite{LSV10}.

\section{Thermodynamics of matter with pressure}
\label{sec:thermodynamics}

Now we address the thermodynamics of the perfect fluid described by the effective Lagrangian in Sec.~\ref{sec1}. Non-dissipative fluids are described by virtue of the \textit{pullback} formalism \cite{1993CQGra..10.2317C,1994CQGra..11..709C,2007LRR....10....1A} through Carter's covariant formulation \cite{1989LNM..1385....1C} in a relativistic effective field theory. In this formulation, an observer is attached to a particular fluid element by introducing a matter space such that its worldline is identified with a unique point in this space. The coordinates of each matter space serve as labels that distinguish fluid element worldlines and remain unchanged throughout the evolution. The matter space coordinates can be considered as scalar fields on spacetime, with a unique map relating them to the spacetime coordinates.
Generally fluids are framed with four scalar fields, namely  $\phi^a$.
In this puzzle three scalar fields, corresponding to $a=1,\,2,\,3$, become fluid comoving coordinates  as they propagate in space, whereas $\phi^0$ is interpreted as an internal time coordinate \cite{2016PhRvD..94l4023B,2016PhRvD..94b5034B,2017arXiv171101961C}.
These scalars can be viewed as St\"uckelberg fields\footnote{This name commonly designates a field that makes explicit a (spontaneously broken) gauge symmetry.} that allow to restore broken diffeomorphisms in four-dimensional spacetimes \cite{2016PhRvD..94l4023B,2003AnPhy.305...96A,2004JHEP...10..076D,2008PhyU...51..759R}. So that, the fluid physical properties are encoded within a set of symmetries of the scalar field action.

We here are interested in cosmological perfect fluids describing the matter sector only. Such a framework may be seen as in Sec.~\ref{sec:action}. Moreover we only deal with the temporal St\"uckelberg field \cite{Matarrese:1984zw,2016PhRvD..94l4023B,2016PhRvD..94b5034B}, hereafter renamed $\varphi$. In other words, providing the cosmological principle, it is licit to take into account that our model is well motivated if one fluid is accounted, say $\phi^0\equiv\varphi$. Without considering the spatial fields implies that the corresponding Lagrangian respects the global shift symmetry
\begin{equation}
\varphi\rightarrow \varphi+c^0\,,
\end{equation}
with $c^0$ an arbitrary constant \cite{2016PhRvD..94l4023B,2016PhRvD..94b5034B}.

The scenario defined in Sec.~\ref{sec:action} turns out to describe a barotropic matter fluid, i.e., its pressure is completely defined by knowing its energy density and viceversa, see Eqs.~(\ref{eq:no11}) and (\ref{eq:no12}).
To provide its thermodynamic interpretation, we choose the particle number density $n$ and the temperature $T$ of the fluid as thermodynamic variables to find correspondence with the field $X$ \cite{2016PhRvD..94l4023B,2016PhRvD..94b5034B}.

To account for perfect fluid thermodynamics we use the first principle, the Gibbs-Duhem relation and the Helmotz free-energy density $f=\rho-Ts$, respectively,
\begin{align}
\label{eq:1pric}
d\rho =\,& T\,ds+\mu\,dn\,, \\
\label{eq:gibbsduhem}
dP =\,& s\,dT +n\,d\mu\,,\\
\label{eq:dhelmotz}
df =\,& \mu\,dn - s\,dT\,.
\end{align}
where $s$ is the entropy density and $\mu$ the chemical potential.
Combining Eqs.~(\ref{eq:1pric})--(\ref{eq:gibbsduhem}) and (\ref{eq:no11})--(\ref{eq:no12}) with the definitions of $f$ we get
\begin{align}
\label{eq:solutions11}
d\left(\mu n\right) =&\, d\left(2X\mathcal{L}_X\right) - d\left(Ts\right)\,,\\
\label{eq:solutions22}
df =&\, d\left(2X\mathcal{L}_X\right) - d\left(K-V^{\rm eff}\right) - d\left(Ts\right)\,.
\end{align}
Keeping in mind that in view of Eq.~(\ref{eq:no5a}) $d\left(K-V^{\rm eff}\right)\equiv d\mathcal{L}$, the above two relations admit as solutions:
\begin{align}
\label{eq:solutions1}
f =\,& -\mathcal{L}\,,\\
\label{eq:solutions2}
s=\,&\sqrt{2X}\mathcal{L}_X\,,\\
\label{eq:solutions3}
T=\,&\sqrt{2X}\,,\\
\label{eq:solutions4}
\mu=\,&0\,.
\end{align}
Eq.~(\ref{eq:solutions1}) fulfills the thermodynamic relations:
\begin{align}
\label{dFdTV}
\frac{\partial\left(fV\right)}{\partial T}\bigg\rvert_V =\,& -\sqrt{2X} \mathcal{L}_X V = -sV\,,\\
\label{dFdVT}
\frac{\partial\left(fV\right)}{\partial V}\bigg\rvert_T =\,& -\mathcal{L}= -\left(K-V^{\rm eff}\right) = -P\,.
\end{align}
From Eq.~\eqref{eq:solutions4} it follows that $df=-sdT$.
The condition $X>0$ and Eq.~\eqref{weaks} automatically define the sign of the entropy density in Eq.~\eqref{eq:solutions2}, i.e. $s\geq0$. Since for expanding systems it has to be $dV>0$ and $dT<0$, necessarily we have that $df>0$. Thence, from Eq.~\eqref{dFdVT} we deduce that $P$ turns out to be \emph{naturally} negative:
\begin{equation}
\label{dfvsP}
 df>0\Leftrightarrow P<0\,.
\end{equation}
Conversely, this is even in agreement with the naive fact that for an the expanding fluid the work is positive. Under our convention of the first principle signs, one thus has: $-PdV>0$ which implies $P<0$.
Eq.~\eqref{dfvsP} implies that \emph{dust-like matter having pressure naturally fixes the sign of $P$ to be negative}. This ensures no need of putting by hand the sign of $P$ inside Einstein's equations to guarantee the universe speed up.

Now we define the densities of internal energy $u$, enthalpy $h$, and Gibbs free-energy $g$ respectively by:
\begin{align}
\label{inendenty}
u = & \rho =\, 2X \mathcal{L}_X - \left(K-V^{\rm eff}\right)\,,\\
\label{enthalpydens}
h = & u + P =\, 2X \mathcal{L}_X\,,\\
\label{gibbsdens}
g = & f + P =\, 0\,.
\end{align}

Invoking the Noether's theorem we notice the global shift symmetry changes the matter Lagrangian density $\mathcal{L}_1$ mostly by a total divergence. We explicitly get:
\begin{align}
\nonumber
&\mathcal{L}_1\left(X^\prime,\varphi^\prime\right) =\mathcal{L}_1\left(\frac{1}{2}\nabla_\alpha\varphi \nabla^\alpha\varphi,\varphi+c^0\right)=\\
\nonumber
&\mathcal{L}_1\left(X,\varphi\right) + c^0\left[\frac{\partial\mathcal{L}_1}{\partial\varphi} - \nabla_\alpha\frac{\partial\mathcal{L}_1}{\partial\left(\nabla_\alpha\varphi\right)}\right] + c^0 \nabla_\alpha\frac{\partial\mathcal{L}_1}{\partial\left(\nabla_\alpha\varphi\right)}=\\
\label{eq:ap5}
&\mathcal{L}_1\left(X,\varphi\right) + c^0 \nabla_\alpha\left(\mathcal{L}_{1,X} \nabla_\alpha\varphi\right)\,,
\end{align}
where, in the second line of Eq.~(\ref{eq:ap5}), the quantity in the brackets identically vanishes in view of the Euler--Lagrange equation.
The conserved current $\mathcal{J}_1^\alpha$ corresponds to the total divergence of Eq.~(\ref{eq:ap5}), i.e.,
\begin{equation}
\label{eq:conscurr1}
\mathcal{J}_1^\alpha = \sqrt{2X} \left(K_X+\lambda Y_X\right) v^\alpha\,.
\end{equation}

At this point it behooves us to discuss how to deal with $V^{\rm eff}$.
Its behavior during the phase transition is not trivial, since it depends on both $\varphi$ and its covariant derivatives via the kinetic term $X$.
This implies that in general, the potential $V^{\rm eff}$ is not invariant under global shift symmetry. On the contrary, $V^{\rm eff}$ is well defined in its minima.
\textit{Before transition} (\textbf{BT}, with $V^{\rm eff}=V_0+\chi \varphi_0^4/4$ at $\varphi=0$) the effective potential is a constant. \textit{After transition} (\textbf{AT}, with $V^{\rm eff}=V_0$ at $\varphi=\varphi_0$), the effective potential is a function of $X$ only, i.e., $V^{\rm eff}\equiv V^{\rm eff}(X)$, which is invariant under global shift symmetry.
For the above reasons, in the following we limit our investigation to BT and AT, where $V^{\rm eff}$ is invariant under shift symmetry. Thus, we do not need to assess the intermediate cases, i.e. during transition.
Therefore, during the BT and the AT phases, the Noether's theorem implies that
\begin{equation}
\label{eq:ap5bis}
\mathcal{L}_2\left(X^\prime\right) = - V^{\rm eff}(X)- c^0 \nabla_\alpha\left(V^{\rm eff}_X \nabla_\alpha\varphi\right)\,,
\end{equation}
where we can define another conserved current from the total divergence of Eq.~(\ref{eq:ap5bis}), i.e.,
\begin{equation}
\label{eq:conscurr1bis}
\mathcal{J}_2^\alpha = -\sqrt{2X} V^{\rm eff}_X v^\alpha\,.
\end{equation}
Hence the total conserved current $\mathcal{J^\alpha}$ is given by combining Eqs.~\eqref{eq:ap5} and \eqref{eq:ap5bis}, i.e.,
\begin{equation}
\label{eq:conscurr}
\mathcal{J}^\alpha \equiv \mathcal{J}_1^\alpha + \mathcal{J}_2^\alpha = \sqrt{2X} \mathcal{L}_X v^\alpha = s^\alpha\,,
\end{equation}
and coincides with the entropy density current $s_\alpha=sv_\alpha$.
Eq.~\eqref{eq:conscurr} simplifies Eq.~(\ref{eq:no5b}) into $\mathcal{L}_\varphi = 0$, \emph{implying that the Lagrangian does not depend upon $\varphi$}. We thus have:
\begin{equation}
\label{eq:1new}
\mathcal{L}\left(\lambda,X,\nu\right) = K(X) - V^{\rm eff}(X) + \lambda Y \left(X,\nu\right)\,.
\end{equation}

By combining Eqs.~(\ref{eq:1pric})--(\ref{eq:gibbsduhem}) we get the Euler relation
 \begin{equation}
\label{eq:euler}
P+\rho = Ts + \mu n\,,
\end{equation}
and recast the energy-momentum tensor as
\begin{equation}
\label{eq:tnesenimp2}
T_{\alpha\beta}=\left(Ts_\alpha +\mu n_\alpha\right)v_\beta + P g_{\alpha\beta}\,,
\end{equation}
where $n_\alpha=nv_\alpha$ is the particle number density current.
The projection of the energy-momentum tensor conservation along $v^\alpha$, i.e., $v^\alpha\nabla^\beta T_{\alpha\beta}=0$, leads to
\begin{equation}
\label{eq:tnesenimp3}
T\nabla^\alpha s_\alpha + \mu \nabla^\alpha n_\alpha = 0\,,
\end{equation}
and by virtue of the existence of $\mathcal{J}^\alpha$, it reduces to
\begin{equation}
\label{eq:tnesenimp4}
\mu \nabla^\alpha n_\alpha = 0\,,
\end{equation}
which represents and identity, since $\mu=0$.
However, one can also safely assume that the particle number density current is also conserved, i.e., $\nabla^\alpha n_\alpha = 0$.

The conservation of the energy-momentum tensor can be recast as the Carter-Lichnerowicz equations \cite{2016PhRvD..94b5034B}
\begin{equation}
\label{eq:cl}
n\mathcal{W}_{\alpha\nu}v^\nu=nT\nabla_\alpha\sigma-\varsigma_\alpha\nabla^\nu n_\nu\,,
\end{equation}
where $\mathcal{W}_{\alpha\nu}=\nabla_\nu \varsigma_\alpha -\nabla_\alpha \varsigma_\nu$ is the vorticity tensor \cite{2013rehy.book.....R}, $\varsigma^\alpha=h/nv^\alpha$ the current of the enthalpy per particle, and $\sigma=s/n$ the entropy per particle.

\noindent Since the $4$-velocity is the derivative of the scalar field $\varphi$ and  $\nabla^\alpha n_\alpha = 0$, we infer from Eq.~(\ref{eq:cl}) that:
\begin{align}
\mathcal{W}_{\alpha\nu}=0\quad&\Rightarrow\quad{\rm the\ fluid\ is\ irrotational}\,,\\
\nabla_\alpha\sigma=0\quad&\Rightarrow\quad{\rm the\ fluid\ is\ isentropic}\,,
\end{align}
respectively from the first and second conditions \cite{2016PhRvD..94b5034B}.

\section{Cosmological perturbations and the role of sound speed}
\label{pert}

In the previous sections we demonstrated that our matter fluid is irrotational and insentropic. We now discuss the cosmological perturbations taking into account our Lagrangian, as in Eq.~\eqref{eq:1new}. We thus unveil additional features characterizing our matter fluid concerning the magnitudes of the pressure $P$ and  the sound speed.

In the conformal Newtonian gauge, in absence of any anisotropic stress, we consider \cite{1999PhLB..458..219G,Gao}
\begin{equation}
\label{pert1}
ds^2=a(\tau)^2\left[\left(1+2\Phi\right)d\tau^2-\left(1-2\Phi\right)dx^2\right]\,,
\end{equation}
where $\Phi$ is the Newtonian potential, $\tau=a(t)t$ the conformal time and $a(t)$ the scale factor.
The first order $\left(0,0\right)$, $\left(0,i\right)$ and $\left(i,j\right)$ components of Einstein's equations in the Friedmann-Robertson-Walker model are respectively,
\begin{align}
\label{pert2}
&\nabla^2\Phi-3\mathcal{H}\left(\Phi^\prime+\mathcal{H}\Phi\right)=4\pi a^2G\delta\rho\,,\\
\label{pert3}
&\nabla_i\left(\Phi^\prime+\mathcal{H}\Phi\right)=4\pi a^2G\left(P+\rho\right) \delta v_i\,,\\
\label{pert4}
&\Phi^{\prime\prime}+3\mathcal{H}\Phi^\prime+\left(2\mathcal{H}^\prime+\mathcal{H}^2\right)\Phi=4\pi a^2G\delta P\,,
\end{align}
where the prime denotes the derivatives with respect to the conformal time and $\mathcal{H}=a^\prime/a$. The perturbation of the 3-velocity $\delta v_i=\nabla_i\delta\varphi/(a\varphi^\prime)$ depends upon the perturbation of the scalar field $\delta\varphi$ which depends also on the spatial coordinates \cite{2014CQGra..31e5006P}. The density perturbations depend on the kinetic term and on the Lagrange multiplier perturbations, respectively $\delta X$ and $\delta\lambda$, whereas the pressure perturbations depend only upon $\delta X$, so that:
\begin{align}
\label{pert5}
\delta\rho=&\,A(X)\delta X+2X\lambda Y_{X\nu}\delta\nu+2XY_X\delta\lambda\,,\\
\label{pert6}
\delta P=&\,B(X)\delta X\,.
\end{align}
To infer the explicit expressions of $A(X)$ and $B(X)$, we discriminate between two regimes \cite{2012CRPhy..13..566M}:
\begin{itemize}
\item[-] BT, when the minimum is at $\varphi=0$ and the potential is $V^{\rm eff}=V_0+\chi \varphi_0^4/4$;
\item[-] AT, when the minimum is at $\varphi=\varphi_0$ and the potential is $V^{\rm eff}=V_0$.
\end{itemize}
Hence, the pressure and density become
\begin{align}
\label{s8}
P(X)&= \left\{
\begin{array}{ll}
K-V_0-\chi \varphi_0^4/4\quad &\quad{\rm (BT)}\\
K-V_0\quad&\quad{\rm (AT)}
\end{array}
\right.,\\
\label{s8bisaaa}
\rho(X)&= \left\{
\begin{array}{ll}
2X\mathcal{L}_X-K+V_0+\chi \varphi_0^4/4\quad &\quad{\rm (BT)}\\
2X\mathcal{L}_X-K+V_0\quad&\quad{\rm (AT)}
\end{array}
\right.,
\end{align}
From the above definitions it follows that
\begin{align}
\label{pert5a}
A(X)&= \left(2X\mathcal{L}_X\right)_X-K_X\,,\\
\label{pert5b}
B(X)&= K_X\,.
\end{align}
Combining Eqs.~\eqref{pert2} and \eqref{pert4} we get
\begin{align}
\nonumber
\Phi^{\prime\prime}&+3\mathcal{H}\left(1+c_{\rm X}^2\right)\Phi^\prime+\left[2\mathcal{H}^\prime+\left(1+3c_{\rm X}^2\right)\mathcal{H}^2\right]\Phi+\\
\label{pert7}
&-c_{\rm X}^2\nabla^2\Phi=4\pi a^2G\left[D(X)\delta\nu+E(X)\delta\lambda\right]\,,
\end{align}
in which
\begin{eqnarray}
c_{\rm X}^2&\equiv& B(X)/A(X)\,,\\
D(X)&\equiv&-2X\lambda Y_{X\nu}c_{\rm X}^2\,,\\
E(X)&\equiv& -2XY_Xc_{\rm X}^2\,.
\end{eqnarray}
The evolution of $\Phi$ in terms of $\rho$ and $\sigma$ perturbations can be written as \cite{1999PhLB..458..219G}
\begin{align}
\nonumber
\Phi^{\prime\prime}&+3\mathcal{H}\left(1+c_{\rm s}^2\right)\Phi^\prime+\left[2\mathcal{H}^\prime+\left(1+3c_{\rm s}^2\right)\mathcal{H}^2\right]\Phi+\\
\label{pert8}
&-c_{\rm s}^2\nabla^2\Phi=4\pi a^2G\zeta\delta\sigma\,,
\end{align}
where
\begin{eqnarray}
c_{\rm s}^2\equiv\partial P/\partial\rho|_\sigma
\end{eqnarray}
is the square of the adiabatic speed of sound and $\zeta\equiv\partial P/\partial\sigma|_\rho$.
From the above considerations, one can define the entropy perturbation shift, $\Delta$, which quantifies how much $\delta P/\delta \rho$ departs from $c_{\rm s}^2$ \cite{2014CQGra..31e5006P}. It can be written as:
\begin{equation}
\label{pert9}
\Delta=\left(\frac{\delta P}{\delta \rho}-c_{\rm s}^2\right)\frac{\delta\rho}{P}=-\frac{D(X)\delta\nu+E(X)\delta\lambda}{P}\,.
\end{equation}
For isentropic fluids (see Sec.~\ref{sec:thermodynamics}), it immediately follows that $\zeta\equiv0$. This is in agreement with our previous outcomes, since from Eq.~\eqref{pert7} we require $Y_X\neq0$,\footnote{That is requested to guarantee the validity of Eq.~\eqref{eq:no9b}.} and so $c_{\rm s}^2\equiv0$ as one assumes $P={\rm const}$ and vice-versa \cite{2009PhRvD..80l3001K,2010PhRvD..82j3535S,2012Ap&SS.338..345L,2014IJMPD..2350012L}.

Taking into account that $P={\rm const}$, we may draw relevant consequences on our fluid temperature.
Indeed, according to Eq.~(\ref{gibbsdens}) we find our fluid to lie on the minimum of the Gibbs energy, i.e. at an equilibrium state.
By combining Eqs.~(\ref{eq:gibbsduhem}), (\ref{eq:dhelmotz}), and (\ref{eq:solutions4}) we get
\begin{equation}
\label{equilibrium}
dg = dP - s\,dT = 0\,.
\end{equation}
Since $P={\rm const}$, one necessarily has $T={\rm const}$ \emph{in the proximity of each minimum of the effective potential}.

Last but not least, it is worth noticing that an isentropic fluid can be even attained from Eq.~\eqref{pert7} by setting $\lambda\rightarrow0$. However, this would represent a particular case for which the Lagrangian term $Y$ has no longer relevance. For the sake of generality, this case is thus excluded into our picture.

\section{Considerations on quantum vacuum energy}
\label{vacuumenergy}

We now analyze in more details the role played by the effective potential $V^{\rm eff}$. In particular, we wonder whether the two possible choices of the off-set $V_0$ provide different physical considerations.
Hence, to alleviate the degeneracy between the two approaches, we need to fix the magnitude associated to $K$. Our target is to bound $K$ in order to heal the fine-tuning issue associated to the cosmological constant $\Lambda$.

We thus explore two possibilities:
\begin{itemize}
\item[1)]{$V_0= -\chi\varphi_0^4/4$, so BT we have $V^{\rm eff}=0$ and hence
\begin{align}
\label{s8bbb}
P_1&= \left\{
\begin{array}{ll}
K\quad &\quad{\rm (BT)}\\
K+\chi \varphi_0^4/4\quad&\quad{\rm (AT)}
\end{array}
\right.,\\
\label{s8bisbbb}
\rho_1&= \left\{
\begin{array}{ll}
2X\lambda Y_X-K\quad &\quad{\rm (BT)}\\
2X\lambda Y_X-K-\chi \varphi_0^4/4\quad&\quad{\rm (AT)}
\end{array}
\right.,
\end{align}
and by virtue of Eq.~\eqref{dfvsP}, then $K<-\chi \varphi_0^4/4$.}
\item[2)]{$V_0=0$, so AT we have $V^{\rm eff}=0$ and hence
\begin{align}
\label{s8ccc}
P_2&= \left\{
\begin{array}{ll}
K-\chi \varphi_0^4/4\quad &\quad{\rm (BT)}\\
K\quad&\quad{\rm (AT)}
\end{array}
\right.,\\
\label{s8bisccc}
\rho_2&= \left\{
\begin{array}{ll}
2X\lambda Y_X-K+\chi \varphi_0^4/4\quad &\quad{\rm (BT)}\\
2X\lambda Y_X-K\quad&\quad{\rm (AT)}
\end{array}
\right.,
\end{align}
and again, by virtue of Eq.~\eqref{dfvsP}, then, $K<0$.}
\end{itemize}
In both cases $K<0$, but with different magnitudes.

\subsubsection{The case $V_0= -\chi\varphi_0^4/4$}

Since $X_\varphi=0$ and $P={\rm const}$, from Eq.~(\ref{eq:no16}) we get that $\eta_\varphi=0$, and, therefore, Eq.~(\ref{eq:no9b}) reduces to
\begin{equation}
\label{eq:no9bb}
\dot{\lambda} = - \theta\lambda\,.
\end{equation}
In the Friedmann-Robertson-Walker spacetime, $\varphi$ is a function of the time only, thus it is easy to demonstrate that $X\equiv\dot{\varphi}^2/2$ and $\theta\equiv 3\dot{a}/a$.
Finally, the solution of Eq.~(\ref{eq:no9bb}) becomes
\begin{equation}
\label{lambdaphia}
\lambda = \lambda_0 a^{-3}\,,
\end{equation}
where $\lambda_0$ is a constant.
Further, in the Friedmann-Robertson-Walker scenario the simplest choice for the (adiabatic) volume may be $\mathcal V=\mathcal V_0a^3$, where  $\mathcal V_0$ the initial volume. Recalling that our  fluid is isentropic, with constant $P$ and $T$, by using Eqs.~(\ref{eq:solutions2}) and (\ref{lambdaphia}), we get
\begin{equation}
\label{isentropic}
s\mathcal{V}=\sqrt{2X}\lambda_0\mathcal V_0 Y_X={\rm const}\,,
\end{equation}
from which it follows that $Y_X=Y_{{\rm BM},X}+Y_{{\rm DM},X}={\rm const}$.

We propose the following assumptions:
\begin{eqnarray}
K_{\rm DM}&\approx&-\chi\varphi_0^4/4\,,\\
K_{\rm BM}&\ll& K_{\rm DM}\,.
\end{eqnarray}
These positions and the fact that $a=(1+z)^{-1}$ (where $z$ is the redshift) allow us to rewrite Eqs.~\eqref{s8bbb}--\eqref{s8bisbbb} as
\begin{align}
\label{s8bbbb}
P_1&\approx \left\{
\begin{array}{ll}
K_{\rm DM}\quad&\quad{\rm (BT)}\\
K_{\rm BM}\quad&\quad{\rm (AT)}
\end{array}
\right.,\\
\label{s8bisbbbb}
\rho_1&\approx \left\{
\begin{array}{ll}
\left(\rho_{\rm DM}+\rho_{\rm BM}\right)\left(1+z\right)^3-K_{\rm DM}\quad &\quad{\rm (BT)}\\
\left(\rho_{\rm DM}+\rho_{\rm BM}\right)\left(1+z\right)^3-K_{\rm BM}\quad&\quad{\rm (AT)}
\end{array}
\right.,
\end{align}
where $\rho_{\rm BM}=2X\lambda_0Y_{{\rm BM},X}$ and $\rho_{\rm DM}=2X\lambda_0Y_{{\rm DM},X}$ are constants.

This mechanism elides the vacuum energy cosmological constant contribution through the use of DM. As $\chi>0$, the sign of $K_{\rm DM}$ is opposite to the vacuum energy term. Hence, from the one hand the DM fluid pushes the universe up to accelerate, while on the other hand vacuum energy provides the opposite contribution in the net pressure.

Then, AT the universe accelerates \emph{because of the presence of a negative baryonic pressure}. This plays the role of \emph{emergent cosmological constant}, which is is negligible with respect to the vacuum energy BT, whereas becomes dominant AT. Since its magnitude is due to the baryon pressure, this alleviates the coincidence problem. In addition the fine-tuning problem is clearly removed because the high value of the predicted vacuum energy density is suppressed and does not enter our framework AT.

\begin{figure}
\centering
\includegraphics[width=\hsize,clip]{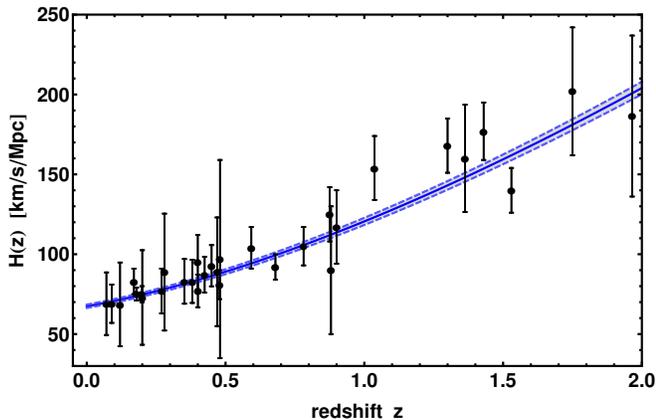}
\caption{$H(z)$ dataset from \cite{2018MNRAS.476.3924C} (black data) compared with the results of our model described by Eq.~\eqref{s8bisbbbb} (solid blue curve). The $1$--$\sigma$ error limits (dashed blue curves) have been obtained by using the best-fit parameters from \cite{2016A&A...594A..13P}.}
\label{fig:1}
\end{figure}

In Fig.~\ref{fig:1} we compare the observational Hubble parameter data (OHD) $H(z)$ (see the black datapoints) with the predictions of our model in Eq.~\eqref{s8bisbbbb}.
The OHD are model-independent measurements of the evolution of the Hubble parameter with redshift from differential age of two galaxies at the same redshift.
The most updated OHD values have been taken from \cite{2018MNRAS.476.3924C}.
From Eq.~\eqref{s8bisbbbb}, we have AT the Hubble parameter can be written as
\begin{equation}
\label{parOL}
H(z)\equiv H_0\sqrt{\frac{2X\lambda_0Y_X}{\rho_{\rm c,0}}\left(1+z\right)^3+\frac{K_{\rm BM}}{\rho_{\rm c,0}}}\,,
\end{equation}
where we can identify the BM+DM density parameter with $\Omega_{\rm m}\equiv 2X\lambda_0Y_X/\rho_{\rm c,0}$ and the dark energy density parameter with $\Omega_\Lambda\equiv -K_{\rm BM}/\rho_{\rm c,0}$. Our predictions can be constrained with the most recent results on the Hubble constant $H_0=(67.74\pm0.46)$~km~s$^{-1}$~Mpc$^{-1}$, the density parameters $\Omega_{\rm m}=0.3089\pm0.0062$ and $\Omega_\Lambda=0.6911\pm0.0062$, and current value of the universe critical density $\rho_{\rm c,0}=(8.62\pm0.12)\times10^{-30}$~g/cm$^3$ obtained by \textit{Planck} \cite{2016A&A...594A..13P}.
These constraints result in the solid blue curve and the $1$--$\sigma$ error limits (the dashed blue curves) shown in Fig.~\ref{fig:1}.

\subsubsection{The case $V_0=0$}

Eqs.~(\ref{eq:no9bb})--(\ref{isentropic}) still hold and retain the same form.
However, in this case the only needed assumption to get the measured cosmological constant is that $K\ll \chi\varphi_0^4/4$.
Therefore we obtain
\begin{align}
\label{s8cccc}
P_2&\approx \left\{
\begin{array}{ll}
-\chi \varphi_0^4/4\quad &\quad{\rm (BT)}\\
K\quad&\quad{\rm (AT)}
\end{array}
\right.,\\
\label{s8biscccc}
\rho_2&= \left\{
\begin{array}{ll}
\left(\rho_{\rm DM}+\rho_{\rm BM}\right)\left(1+z\right)^3+\chi \varphi_0^4/4\ &\ {\rm (BT)}\\
\left(\rho_{\rm DM}+\rho_{\rm BM}\right)\left(1+z\right)^3-K\ &\ {\rm (AT)}
\end{array}
\right.,
\end{align}
where the BM can be considered even pressureless.

In this case the vacuum energy density cancels without the effect of any matter component.
This occurrence is due to the discontinuity of the effective potential introduced by the phase transition only.
The emergent cosmological constant appears soon after the transition as related to the DM sector of the universe and holds the \textit{ad hoc} value to justify the observed acceleration of the universe.
Therefore, this case still suffers from the coincidence problem, which affects the $\Lambda$CDM model.
Moreover, differently from the previous case and in analogy with the concordance model, the baryons do not play a significant role in speeding up the universe. Indeed, they can be viewed as pressureless particles.

\section{Temperature and mass of the DM candidate}
\label{DMparticle}

As discussed above, the $V_0= -\chi\varphi_0^4/4$ case is preferred over $V_0=0$, to avoid discontinuities in the pressure contribution. In so doing, one may break the degeneracy between the two approaches, choosing the case $V_0= -\chi\varphi_0^4/4$ which corresponds to a dark fluid defined by matter with pressure.
Thus, limiting on  $V_0= -\chi\varphi_0^4/4$ we draw in the thermal universe the bounds over the DM constituent as particle candidate for DM enabling the process for that the DM pressure elides the vacuum energy contribution.\footnote{
For the sake of clearness, we hereafter restore the usual physical constants, previously set to $1$.}

The energy and number densities, together with the pressure of each particles having mass $m$, momentum $p$ and equilibrium temperature $T$, can be computed as \cite{2008cosm.book.....W}
\begin{align}
\label{fd_endens}
\epsilon=&\,g\frac{(k_{\rm B}T)^4}{2\pi^2\hbar^3c^3}\int^{\infty}_{0}\frac{\xi^2\sqrt{\xi^2+A^2}}{e^{\sqrt{\xi^2+A^2}}\pm1}d\xi\,,\\
\label{fd_partdens}
n=&\,g\frac{(k_{\rm B}T)^3}{2\pi^2\hbar^3c^3}\int^{\infty}_{0}\frac{\xi^2}{e^{\sqrt{\xi^2+A^2}}\pm1}d\xi\,,\\
\label{fd_pressdens}
P=&\,g\frac{(k_{\rm B}T)^4}{2\pi^2\hbar^3c^3}\int^{\infty}_{0}\frac{\xi^4}{3\sqrt{\xi^2+A^2}}\frac{d\xi}{e^{\sqrt{\xi^2+A^2}}\pm1}\,,
\end{align}
where $\xi=pc/(k_{\rm B}T)$, $A=mc^2/(k_{\rm B}T)$, $g=2s+1$ is the spin $s$ degeneracy parameter, $c$ the speed of light, $\hbar$ the reduced Planck constant and $k_{\rm B}$ the Boltzmann constant. Here the choice ``$\pm$'' distinguishes fermions and bosons, respectively.
The entropy density is simply given by
\begin{equation}
\label{fd_entrdens}
s=\frac{g k_{\rm B}^4T^3}{6\pi^2\hbar^3c^3}\int^{\infty}_{0}\frac{4\xi^2+3A^2}{\sqrt{\xi^2+A^2}}\frac{\xi^2d\xi}{e^{\sqrt{\xi^2+A^2}}\pm1}\,.
\end{equation}
We focus on bosons since $\mathcal L_1$  has been written for bosons only. The DM constituents are in our picture bosons that at early times behave as relativistic particles ($mc^2\ll k_{\rm B}T$), in thermal equilibrium.
The energy density of all bosons (b) and fermions (f) species comes by summing up Eq.~\eqref{fd_endens} for each of them, i.e.,
\begin{equation}
\label{endenstot2}
\epsilon_{\rm BT}=g_{*}\frac{\pi^2(k_{\rm B}T_{\rm p})^4}{30(\hbar c)^3}\,,
\end{equation}
where $g_{*}$ is the sum of the standard term $g_{*}^{\rm ST}=\sum_b g_b+\frac{7}{8}\sum_f g_f\approx 106.75$ \cite{2008cosm.book.....W} and our DM particle term $g_{\rm DM}=2s_{\rm DM}+1$ with spin $s_{\rm DM}$.
Independently from the offset on $V_0$, the BT total energy density in Eqs.~\eqref{s8bbbb}--\eqref{s8biscccc} is given by
\begin{equation}
\label{endenstot}
\epsilon_{\rm BT}=\left[\Omega_{\rm r}\left(\frac{T_{\rm p}}{T_0}\right)^4 + \Omega_{\rm m}\left(\frac{T_{\rm p}}{T_0}\right)^3 + \Omega_\Lambda \right]\epsilon_{\rm c} + \epsilon_{\rm v}\,,
\end{equation}
where $\Omega_{\rm r}=(9.16\pm0.19)\times10^{-5}$ is the radiation density parameter \cite{2016A&A...594A..13P}, and $\epsilon_{\rm c}=\rho_{\rm c}c^2$, where $\rho_{\rm c}=3H^2/(8\pi G)$ is the universe critical density, in which $H$ is the Hubble parameter and $G$ the gravitational constant.
As already stated in Sec.~\ref{vacuumenergy}, the current value of the universe critical density is $\rho_{\rm c,0}=(8.62\pm0.12)\times10^{-30}$~g/cm$^3$ \cite{2016A&A...594A..13P}.
With respect to Eqs.~\eqref{s8bbbb}--\eqref{s8biscccc}, we include the radiation, which is not  negligible at early times, and use the relation $T_{\rm p}/T_0=(1+z)$, in which $T_{\rm p}$ is the cosmic plasma temperature and $T_0=2.725$~K the current \textit{Cosmic Microwave Background} temperature.\footnote{AT till today the universe is in equilibrium. Therefore, as in Eq.~\eqref{isentropic}, the entropy conservation implies that $s\propto a^{-3}$ and $T\propto a^{-1}$, as follows from Eq.~\eqref{fd_entrdens}, whereas during the phase transition the temperature has a constant value $T_{\rm p}$.}
Finally, $\epsilon_{\rm v}=7.74\times10^{46}$~erg/cm$^3$ is the vacuum energy density.
By equating Eq.~\eqref{endenstot} and Eq.~\eqref{endenstot2} and solving numerically, we get the plasma temperature
\begin{equation}
T_{\rm p}=(6.6559\pm0.0019)\times10^{14}h(s_{\rm DM})~K\,,
\end{equation}
where $h(0)=1$, $h(1)=0.995$, and $h(2)=0.991$.

\begin{figure}
\centering
\includegraphics[width=\hsize,clip]{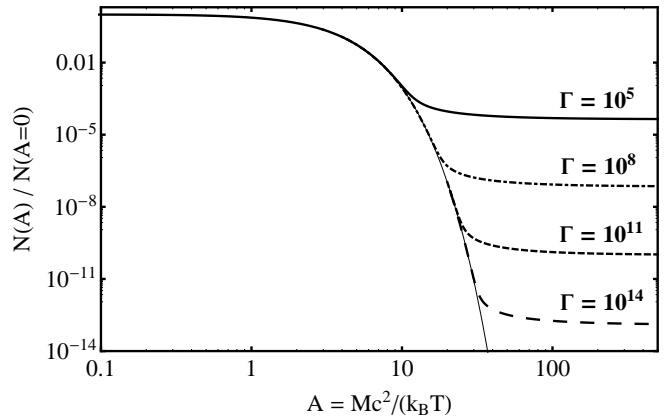}
\caption{Plot of $N(A)$ for $\Gamma=10^5$,\,$10^8$,\,$10^{11}$,\,$10^{14}$. The freeze-out occurs at $A_{\rm f}=11$,\,$17$,\,$25$,\,$32$, respectively.}
\label{fig:2}
\end{figure}

The primordial DM interactions can be viewed as the annihilation of a heavier DM particle $Q$ and its antiparticle $\bar{Q}$, both with masses $M$, to produce two lighter particles $q$ and $\bar{q}$.
Assuming no initial asymmetry between the particles $Q$ and $\bar{Q}$, their comoving density must be the same, i.e., $n_{Q}\equiv n_{\bar{Q}}\equiv n$; on the other hand $q$ and $\bar{q}$ are tightly coupled to the cosmic plasma.
Therefore the Boltzmann equation for the evolution of $n$ writes as
\begin{equation}
\label{NQ}
\frac{1}{a^3}\frac{d \left(a^3n\right)}{dt}=-\langle \varkappa v\rangle \left(n^2-n_{\rm eq}^2\right)\,,
\end{equation}
where $n_{\rm eq}$ is the equilibrium number density and $\langle \varkappa v\rangle$ the thermally averaged cross-section.
From the entropy conservation we write the number density as an adimensional quantity $N=n k_{\rm B}/s$.
Then, we note that the comoving time $t$ is related to $A$ by: $dA=HA\,dt$.
Before neutrino decoupling, the entropy density degeneracy parameters is $g_{\rm s}^{*}\equiv g_{*}$ \cite{2008cosm.book.....W}, therefore the total entropy density is given by $s=2\pi^2 k_{\rm B}^4 g_{*} T^3/[45(\hbar c)^3]$.
From the identity $\epsilon_{\rm BT}\equiv\rho_{\rm c}c^2$, we obtain
\begin{equation}
\label{NQ4}
H\equiv\left(\frac{\dot a}{a}\right)=\sqrt{\frac{4\pi^3c^3g_{*}G}{45\hbar^3}}\left(\frac{M}{A}\right)^2\,.
\end{equation}
From the above definitions, we can recast the Boltzmann equation to obtain a Riccati-like equation
\begin{equation}
\label{NQ3}
\frac{d N}{dA}=-\frac{\Gamma}{A^2}\left(N^2-N_{\rm eq}^2\right)\,,
\end{equation}
where we defined the interaction rate
\begin{equation}
\label{Gamma}
\Gamma\equiv \sqrt{\frac{g_{*}\pi c^3}{45G\hbar^3}} \langle\varkappa v\rangle M\,.
\end{equation}
Fig.~\ref{fig:2}, shows $N(A)$ for $\Gamma=10^5$, $10^8$, $10^{11}$, and $10^{14}$. The value of the DM relic abundance is given by
\begin{equation}
N_\infty\approx A_{\rm f}/\Gamma\,,
\end{equation}
where $A_{\rm f}$ marks the transition from the relativistic regime to the non-relativistic one.
For the above wide range of $\Gamma$, we can safely assume that the non-relativistic regime is attained for $A_{\rm f}=10$--$30$.
We assume that at this stage the temperature is approximately the above equilibrium temperature $T_{\rm p}$.
For this choice the DM particle mass stays approximately in the range of values $0.5\lesssim M{\rm c^2/ TeV} \lesssim 1.7$, in agreement with the most recent predictions over the WIMPs \cite{2002JHEP...12..034K,PhysRevLett.84.5699,2002PhLB..545...43B}.
The precise values depend upon on the value of $h$, which is quite insensitive to $s_{\rm DM}$, as summarized in Tab.~\ref{tab:table1}.

\begin{table}
\caption{\label{tab:table1}The mass range of the DM boson particle candidate depending on the spin particle. Columns list DM spin $s_{\rm DM}$, the function $h(s_{\rm DM})$, and the mass range of the DM particle.}
\begin{ruledtabular}
\begin{tabular}{lcr}
$s_{\rm DM}$	&	h($s_{\rm DM}$)	&	M (TeV)\\
\hline
0				& 	1.000			& 	0.574--1.723\\
1				&	0.995			&	0.572--1.715\\
2				&	0.991			&	0.569--1.708
\end{tabular}
\end{ruledtabular}
\end{table}

Using the above definitions, we now relate the freeze-out abundance of DM relics to its density today, i.e., $\rho_{\rm Q,0}=N_\infty s_0 M$. The DM density parameter is
\begin{equation}
\label{relic}
\Omega_{\rm Q}=\frac{\rho_{\rm Q,0}}{\rho_{\rm c,0}}=\frac{16g^{*}_{\rm s,0}}{3H_0^2}\sqrt{\frac{G^3\pi^5\hbar^3}{45g_{*}c^3}}\left(\frac{k_{\rm B}T_0}{\hbar c}\right)^3\frac{A_{\rm f}}{\langle \varkappa v\rangle}\,,
\end{equation}
where $g^{*}_{\rm s,0}=3.91$, and $A_{\rm f}=10$--$30$.
Within the proposed case $V_0= -\chi\varphi_0^4/4$, by looking at Eq.~\eqref{s8bisbbbb} we can impose $\Omega_{\rm Q}\equiv\Omega_{\rm dm}=0.2589\pm0.0057$ \cite{2016A&A...594A..13P} in Eq.~\eqref{relic}.
This position provides a range of values for the thermally averaged cross-section $0.81\leq\langle \varkappa v\rangle/(10^{-26}{\rm cm}^3{\rm s}^{-1})\leq2.42$\,.\footnote{For completeness, one may also deal with the case $V_0=0$, which corresponds to the $\Lambda$CDM case.
By looking at Eq.~\eqref{s8biscccc}, this time we are forced to impose $\Omega_{\rm Q}\equiv\Omega_{\rm dm}+\Omega_\Lambda$ in Eq.~\eqref{relic}. This position gives as range $2.20\leq\langle \varkappa v\rangle/(10^{-27}{\rm cm}^3{\rm s}^{-1})\leq6.61$\,. This case however, albeit degenerating with the previous one, is not favored for the requests we made in the previous section.}

\section{Predictions of our paradigm}
\label{predictions}

We here sum up the main results of our paradigm. We revise the concordance model, assuming the most general Lagrangian for matter with pressure. To do so, we consider a transition phase induced by the effective potential of a vacuum energy cosmological constant, with a mechanism in which the DM pressure elides the vacuum energy pressure itself. So that we obtain:
\begin{align}
&P={\rm const}\ {\rm (always)}\,\Rightarrow\,c_{\rm s}^{\rm DM}\equiv c_{\rm s}^{\rm BM}\equiv0\,,\\
&P<0\ {\rm (from\ thermodynamics)}\,,\\
&T={\rm const}\ {\rm (during\ the\ transition)}\,,\\
&P_{\rm DM}\gg P_{\rm BM}\ ,\ P_{\rm DM}\approx\epsilon_{\rm v}\,,\\
&\rho_\Lambda\equiv P_{\rm DM}\ {\rm (BT)}\,,\\
&\rho_\Lambda\equiv P_{\rm BM}\ {\rm (AT)}\,,\\
&0.5\lesssim M{\rm c^2/ TeV} \lesssim1.7 \ {\rm (Cold\ Dark\ Matter)}\,,\\
&0.81\leq\langle \varkappa v\rangle/(10^{-26}{\rm cm}^3{\rm s}^{-1})\leq2.42\,.
\end{align}
Hence, in our scheme there exists only \emph{one perfect, irrotational, and isentropic fluid}, composed of BM and DM. $\Lambda$ is coupled with the matter.
The thermodynamics of such a fluid \emph{naturally} suggests an emergent negative pressure.
The effective potential $V^{\rm eff}$ induces a transition phase during which the quantum vacuum energy density mutually cancels with the DM pressure.
Soon after the transition the emergent cosmological constant is given by the (negative) pressure of baryons.
This overcomes the fine-tuning problem between the predicted and observed values of $\Lambda$ and the coincidence problem, due to the fact that it is the matter which induces the effective cosmological constant at late times and, therefore, it is natural that their magnitudes are extremely close today.
The model mimes the $\Lambda$CDM effects, without departing from observations made at both late and early stages of universe's evolution \cite{2012PhRvD..86l3516A,2016A&A...594A..13P}.

As principal responsible for DM in the universe, our predictions on the mass constituents, i.e. $0.5\lesssim M\lesssim1.7$ TeV, leave open the possibility to detect in laboratory additional heavier bosons, e.g. for example additional $Z^\prime$ or $W^\prime$ bosons or Leptoquarks as potentially predicted by extensions of the particle standard model.

\section{Final outlooks and perspectives}
\label{conclusions}

In this work, we proposed an alternative model to the standard $\Lambda$CDM paradigm. We assumed the existence of a single fluid composed by matter only, i.e. baryons and cold DM. The fluid pushes the universe up, canceling the quantum contribution due to the cosmological constant through the assumption that matter shows a non-vanishing pressure. In particular, we proposed that both DM and BM are collisional, through a generalized scalar field $\varphi$  representation of the matter fluid Lagrangian $\mathcal{L}_1$ depending upon a kinetic term, $X$, and a Lagrange multiplier, $\lambda$.
We even included a potential, $V^{\rm eff}$, which models the coupling with the standard cosmological constant and induces a phase transition. We described the thermodynamics of our matter fluid, showing that it is \emph{perfect, irrotational, and isentropic}. Moreover, we demonstrated that the positiveness of the Helmotz energy \emph{naturally} suggests a negative pressure.
We showed the existence of a Noether current due to the shift symmetry, which coincided with the entropy density current $s^\alpha$, making the Lagrangian independent from $\varphi$. Thus,  we assumed a homogeneous and isotropic space-time to investigate small perturbations and we found that the adiabatic sound speed naturally vanishes, leading to a constant pressure, but with an evolving energy density, differently from the standard $\Lambda$CDM model. To this end, we mostly analyzed the role of the effective potential, $V^{\rm eff}$. To do so, we managed the off-set $V_0$ by analyzing two possibilities: $V_0= -\chi\varphi_0^4/4$ and $V_0=0$.

In the first case ($V_0= -\chi\varphi_0^4/4$), the effective potential induced a phase transition during which the quantum vacuum energy density mutually cancels with the DM pressure. This mechanism has consequences even as the transition stops. Indeed, soon AT the emergent cosmological constant, able to accelerate the universe today, is given by the (negative) pressure of BM. This achievement overcomes both the fine-tuning and the coincidence problem. The fine-tuning problem is overcome since the contributions due to the vacuum energy is canceled out through DM. The coincidence problem is healed since it is the matter which induces the effective cosmological constant at late times. Thus, it is natural to presume that their magnitudes are extremely close today.

In the second case ($V_0=0$), the DM does not play an active role in erasing the quantum field vacuum energy density and BM can be viewed as pressureless. Hence, this landscape does not offer solutions for the fine-tuning and the coincidence caveats. As a consequence it degenerates with the previous case and can be identified with the standard $\Lambda$CDM paradigm. The so-obtained dark fluid is thus mimed by a matter fluid with pressure in which the minimum favors the first case. Further, both cases manifested constant pressure and constant Gibbs free-energy during the universe evolution, with $T={\rm const}$ during transition as naturally expected for first order phase transition.

In addition, we related the predictions of our model to observations by directly comparing the energy density of the cosmic plasma with the one described by our matter fluid. The temperature at which the transition occurred is in quite good agreement with early-time temperatures of hot plasma. Afterwards, from the study of the DM relic abundance for the preferred case with $V_0= -\chi\varphi_0^4/4$, we posed stringent limits on the mass, $0.5\lesssim M{\rm c^2/ TeV} \lesssim1.7$, and the thermally averaged cross-section, $0.81\leq\langle \varkappa v\rangle/(10^{-26}{\rm cm}^3{\rm s}^{-1})\leq2.42$, of the DM particle candidate. These estimates are quite independent from the spin of DM particles.

In future works, we will study inflationary scenarios, naturally arising from Eqs.~\eqref{s8bbbb}--\eqref{s8bisbbbb}, through our hypothesis of matter with non-vanishing pressure. We will better analyze also additional symmetries of our Lagrangians and we will put more stringent constraints on the DM particle, bounding the cross-section from current DM experiments.

\begin{acknowledgments}
O.~L. thanks Danilo Babusci, Stefano Bellucci, Salvatore Capozziello, Raffaele Marotta, Hernando Quevedo, Luigi Rosa and Patrizia Vitale for their suggestions. He is also grateful to Peter K.~S. Dunsby for the discussions on the topic of adiabatic fluids in cosmology, made at the University of Cape Town. M.~M thanks Enrico Nardi for enlightening discussions about the topic of dark matter relic abundance.
\end{acknowledgments}

\appendix
\section{Derivation of the equations in Sec.~\ref{sec:action}}
\label{appA}

From the variation of the action in Sec.~\ref{sec:action} we get
\begin{widetext}
\begin{align}
\nonumber
\delta S = &\int \left[ \left(K_\varphi - V^{\rm eff}_\varphi + \lambda Y_\nu\nu_\varphi \right) \delta\varphi + \mathcal{L}_X \nabla_\alpha\varphi \nabla^\alpha \delta \varphi + Y \delta\lambda + \frac{1}{2} g_{\alpha\beta} \left(\mathcal{L}_X \nabla_\alpha\varphi \nabla^\alpha \varphi - K + V^{\rm eff} - \lambda Y \right) \delta g^{\alpha\beta}\right] \sqrt{-g}{\rm d}^4x =\\
\label{eq:ap1}
= & \int \left\{ \left[ \mathcal{L}_\varphi - \nabla_\alpha \left(\mathcal{L}_X \nabla^\alpha \varphi \right) \right] \delta\varphi + Y \delta\lambda + \frac{1}{2} g_{\alpha\beta} \left( \mathcal{L}_X \nabla_\alpha\varphi \nabla^\alpha \varphi - K + V^{\rm eff} - \lambda Y \right) \delta g^{\alpha\beta} \right\} \sqrt{-g}{\rm d}^4x\,.
\end{align}
\end{widetext}
The variations with respect to $\lambda$, $\varphi$ and $g_{\alpha\beta}$, respectively, lead to Eqs.~(\ref{eq:no5a}), (\ref{eq:no5b}), and (\ref{eq:no6}).

Eq.~(\ref{eq:no13}), instead, is obtained trought simple calculations including the vanishing acceleration in Eq.~(\ref{eq:no8})
\begin{align}
\nonumber
\nabla_\alpha T^{\alpha\beta} =\,& \dot{\rho} v^\beta + \nabla_\alpha P v^\alpha v^\beta + \theta \left( \rho + P \right) v^\beta - \nabla^\beta P =\\
\label{eq:ap2}
=\,& \left[ \dot{\rho} + \theta \left( \rho + P \right) \right] v^\beta\,,
\end{align}
and the expansion $\theta$ in Eq.~(\ref{eq:no15}) is obtained as
\begin{equation}
\label{eq:ap3}
\theta = \frac{\nabla_\alpha \nabla^\alpha \varphi}{\sqrt{2X}} - \frac{X_\varphi}{\sqrt{2X}}\frac{\nabla_\alpha \varphi \nabla^\alpha \varphi}{2X} = \frac{\nabla_\alpha \nabla^\alpha \varphi - X_\varphi}{\sqrt{2X}}\,.
\end{equation}

By using Eqs.~(\ref{eq:no5a}) and (\ref{eq:no16}), we can recast Eq.~\eqref{eq:no5b} to obtain Eq.~(\ref{eq:no9b})
\begin{widetext}
\begin{align}
\nonumber
&K_\varphi - V^{\rm eff}_\varphi + \lambda Y_\nu \nu_\varphi - \nabla_\alpha \left( K_X - V^{\rm eff}_X + \lambda Y_X \right)\nabla^\alpha \varphi - \left( K_X - V^{\rm eff}_X + \lambda Y_X \right)\nabla_\alpha\nabla^\alpha \varphi =\\
\nonumber
=\,&\mathcal{L}_\varphi- \left[ K_{XX}X_\varphi + K_{X\varphi} - V^{\rm eff}_{XX}X_\varphi - V^{\rm eff}_{X\varphi} + \lambda\left(Y_{XX}X_\varphi+Y_{X\nu}\nu_\varphi\right)\right]\nabla_\alpha\varphi\nabla^\alpha\varphi - \mathcal{L}_X \nabla_\alpha\nabla^\alpha \varphi - \nabla_\alpha\lambda Y_X \nabla^\alpha \varphi =\\
\label{eq:ap4}
=\,&\mathcal{L}_\varphi - 2X \left(\mathcal{L}_{XX} X_\varphi + \mathcal{L}_{X\varphi} \right) - \left(\sqrt{2X}\theta+X_\varphi\right) \mathcal{L}_X - \sqrt{2X}\dot{\lambda} Y_X = -\eta_\varphi - \frac{\theta}{\sqrt{2X}}\left(\rho+P\right) - \sqrt{2X}\dot{\lambda} Y_X =0\,.
\end{align}
\end{widetext}

\end{document}